\documentclass[12pt]{iopart}
\usepackage{ae}
\usepackage{graphicx}
\usepackage{color}
\usepackage{wasysym}

\usepackage{hyperref}
\hypersetup{
		colorlinks=true,
		pdfstartview=FitV,
		linkcolor=blue,
		citecolor=blue,
		urlcolor=blue,
		pdfauthor={STAHL Adam, stahla@chalmers.se},
		plainpages=false,
	}

\newcommand{\pder}[2]{\ensuremath{\frac{\partial #1}{\partial #2}}}

\newcommand{\reference}[1]{\ensuremath{\tilde{#1}}}
\newcommand{\normalized}[1]{\ensuremath{\bar{#1}}}
\renewcommand{\t}[1]{\mbox{\scriptsize #1}}
\newcommand{\rve}{\reference{v}_{\t{e}}}
\newcommand{\ry}{y}
\newcommand{\rx}{x}
\newcommand{\rT}{\reference{T}}
\newcommand{\rn}{\reference{n}}
\newcommand{\rdelta}{\delta}
\newcommand{\rnuee}{\reference{\nu}_{\t{ee}}}
\newcommand{\rlnL}{\ln\reference{\Lambda}}
\newcommand{\bve}{\normalized{v}_{\t{e}}}
\newcommand{\nuee}{\nu_{\t{ee}}}
\newcommand{\bnuee}{\normalized{\nu}_{ee}}

\newcommand{\CODE}{\textsc{CODE}}
\newcommand{\zeff}{Z_{\mbox{\scriptsize eff}}}
\newcommand{\power}[1]{\!\cdot\! 10^{#1}}

\newcommand{\EH}{\ensuremath{\hat{E}}}
\newcommand{\tH}{\ensuremath{\hat{t}}}
\newcommand{\Ctp}{C^{\t{tp}}}	 
\newcommand{\Cfp}{C^{\t{fp}}}
\newcommand{\Cnl}{C^{\t{nl}}}
\newcommand{\ava}{\mbox{\scriptsize }}
\newcommand{\Ch}{\t{Ch}}
\newcommand{\RP}{\t{RP}}
\newcommand{\Dr}{\t{D}}
\newcommand{\tin}{\t{in}}
\newcommand{\tmax}{\t{max}}

\newcommand{\g}{\gamma}
\newcommand{\gin}{\gamma_{\tin}}
\newcommand{\lesssim}{\apprle}
\newcommand{\gtrsim}{\apprge}
\renewcommand{\d}{\mbox{d}}

\newcommand{\Eq}[1]{Eq.~(\ref{#1})}
\newcommand{\Fig}[1]{Fig.~\ref{#1}}
\newcommand{\Ref}[1]{Ref.~\cite{#1}}

\newcommand{\blue}[1]{#1}

\begin{document}

\title[Kinetic modelling of runaway electrons in dynamic scenarios]{Kinetic modelling of runaway electrons in dynamic scenarios}

\author{
		A Stahl$^{1}$,  
		O Embr\'eus$^{1}$, 		 
		G Papp$^{2}$,
		M Landreman$^{3}$ and
		T F\"ul\"op$^{1}$
		}

\address{
		$^{1}$ Department of Physics, Chalmers University of Technology, G\"oteborg,  Sweden
		}
\address{
		$^{2}$ Max Planck Institute for Plasma Physics, Garching, Germany
		}
\address{
		$^{3}$ University of Maryland, College Park, MD, USA
		}
\ead{stahla@chalmers.se}

\begin{abstract}
Improved understanding of runaway-electron formation and decay processes are of prime interest for the safe operation of large tokamaks, and the dynamics of the runaway electrons during dynamical scenarios such as disruptions are of particular concern. In this paper, we present kinetic modelling of scenarios with time-dependent plasma parameters; in particular, we investigate hot-tail runaway generation during a rapid drop in plasma temperature. With the goal of studying runaway-electron generation with a self-consistent electric-field evolution, we also discuss the implementation of a collision operator that conserves momentum and energy and demonstrate its properties. An operator for avalanche runaway-electron generation, which takes the energy dependence of the scattering cross section and the runaway distribution into account, is investigated. We show that the \blue{simplified} avalanche model of Rosenbluth and Putvinskii (1997 \emph{Nucl. Fusion} {\bf 37} 1355) can give inaccurate results for the avalanche growth rate (either lower or higher) for many parameters, especially when the average runaway energy is modest, such as during the initial phase of the avalanche multiplication. The developments presented pave the way for improved modelling of runaway-electron dynamics during disruptions or other dynamic events.
\end{abstract}


\section{Introduction}
Runaway electrons, a phenomenon made possible by the decrease of the collisional friction with particle energy \cite{Dreicer1,Dreicer2}, are common in plasmas in the presence of strong external electric fields or changing currents. The tightly focused beam of highly relativistic particles can be a serious threat to the first wall of a fusion reactor, due to the possibility of localized melting or halo-current generation \cite{ITER_DMS}. In the quest for avoidance or mitigation of the harmful effects of runaway-electron losses, a greater understanding of the runaway-electron phenomenon is required \cite{Boozer}. Improved knowledge of runaway-electron formation mechanisms, dynamics and characteristics will benefit the fusion community and contribute to a stable and reliable operation of reactor-scale tokamaks.

Kinetic simulation is the most accurate and useful method for investigating runaway-electron dynamics, and we recently developed a new tool called \CODE{} (COllisional Distribution of Electrons \cite{CODE}) for fast and detailed study of these processes. \CODE{} solves the spatially homogeneous kinetic equation for electrons in 2-D momentum space, including electric-field acceleration, collisions, avalanche runaway generation and synchrotron-radiation-reaction losses \cite{CODE,PRL,Bump_ana}. In \CODE{}, momentum space is discretized using finite differences in momentum and a Legendre-mode decomposition in pitch-angle cosine. Often, the time evolution of the distribution is the desired output, but a (quasi-)steady-state solution can also be efficiently obtained through the inversion of a single sparse system (in the absence of an avalanche source). \CODE{} has been used to study the spectrum of the synchrotron radiation emitted by runaways \cite{CODE}, the corresponding influence of the emission on the distribution function \cite{PRL,Bump_ana,Bump_num}, and the factors influencing the critical electric field for runaway-electron generation \cite{PRL,Paz-Soldan}. 

In this paper we describe improvements to \CODE{} which enable us to investigate the effect of hot-tail runaway generation on the distribution (Section~\ref{sec:td}). This process can be the dominant mechanism in rapidly cooling plasmas. We also discuss the implementation of a full linearized collision operator, and demonstrate its conservation properties (Section \ref{sec:coll_op}). The use of this operator is necessary in cases where the correct plasma conductivity is required, and our implementation indeed reproduces the Spitzer conductivity \cite{SpitzerHarm} for weak electric fields. In addition, an improved model for the large-angle (knock-on) Coulomb collisions leading to avalanche multiplication of the runaway population \cite{Chiu}, is described in Section \ref{sec:knock-on}. This model takes the energy dependence of the runaway distribution into account, and uses the complete energy-dependent M\o ller scattering cross section \cite{Moller}. We find that its use can in some cases lead to significant modifications to the avalanche growth rate, compared to \blue{the more simplified model of Rosenbluth \& Putvinskii} \cite{RosPut}. 

The improvements described in this work enable the detailed study of runaway processes in dynamic situations such as disruptions, and the conservative collision operator makes self-consistent calculations of the runaway population and current evolution in such scenarios feasible \cite{PappEPS15}.

\section{Time-dependent plasma parameters}
\label{sec:td}
To be able to investigate the behavior of the electron population in dynamic scenarios such as disruptions or sawtooth crashes, it is necessary to follow the distribution function as the plasma parameters change. To this end, \CODE{} has been modified to handle time-dependent background-plasma parameters. Since the kinetic equation is treated in linearized form, the actual temperature and density of the distribution are determined by the background Maxwellian used in the formulation of the collision operator. This allows for a scheme where the kinetic equation is normalized to a \emph{reference} temperature $\rT$ and number density $\rn$, so that the discretized equation can be expressed on a fixed \emph{reference grid} in momentum space (throughout this paper, we will use a tilde to denote a reference quantity). By changing the properties of the Maxwellian equilibrium around which the collision operator is linearized, the evolution of the plasma parameters can be modelled on the reference grid without the need for repeated interpolation of the distribution function to new grids. 

Analogously to \Ref{CODE}, the kinetic equation in 2D momentum space for the electron distribution function $f$ experiencing an electric field $E$ (parallel to the magnetic field) and collisions, can be expressed as
\begin{equation}
	\pder{F}{\tH} + 
	\EH  \left(  \xi\pder{F}{\ry} + \frac{1-\xi^{2}}{\ry}\pder{F}{\xi}  \right)
	=\hat{C}\left\{ F\right\} +\hat{S}_{\ava}\left\{ F\right\}.
\label{eq:kinetic}
\end{equation}
Here we have introduced a convenient normalized momentum $\ry \!=\! \gamma v/\rve$, where $\rve\! =\! \sqrt{2\rT/m}$ is the reference electron thermal speed, and the cosine of the pitch angle $\xi=\ry_\|/\ry$. Using  $\kappa\!=\!m^{3}\rve^{3}\pi^{3/2}/\rn$, we have also defined the distribution function $F\!=\!F(y,\xi)\!=\!\kappa f$ (normalized so that $F(\ry\!=\!0)=1$ for a Maxwellian with $T\!=\!\rT$ and $n\!=\!\rn$), time $\tH \!=\! \rnuee t$, and electric field $\EH\!=\!-eE/m\rve\rnuee$, as well as the normalized operators $\hat{C} \!=\! C\,\kappa/\rnuee$ and $\hat{S}_{\ava}\!=\!S_{\ava}\kappa/\rnuee$, with $\rnuee = 16\sqrt{\pi}e^4\rn\rlnL / 3m^2\rve^3$ the reference electron thermal \blue{collision frequency}, $-e$, $m$ and $v$ the charge, rest mass and speed of the electron, and $\gamma$ the relativistic mass factor. Note that $|\EH| = (3\sqrt{\pi}/2) E/E_{\t{D}}$, with $E_{\t{D}}$ the Dreicer field \cite{Dreicer1}. $C$ is the Fokker-Planck collision operator and $S_{\ava}$ an operator describing close (large-angle) Coulomb collisions. These operators will be discussed more thoroughly in Sections~\ref{sec:coll_op} and \ref{sec:knock-on}, respectively; for now we just state the formulation of the collision operator employed in \Ref{CODE} using the normalizations above:
\begin{equation}
	\hat{C}^{\mbox{\scriptsize tp}} = c_C \bve^3 \ry^{-2} 
	\left(    
		\pder{}{\ry}
			\left[ 
				\ry^2\Psi\,\left( \frac{1}{\rx} \pder{}{\ry} + \frac{2}{\bve^2} \right)\!F
			\right]
		+ \frac{c_\xi}{2\rx} \pder{}{\xi}(1-\xi^2)\pder{F}{\xi}
	\right).\label{eq:C}
\end{equation}
\blue{The superscript $\mbox{\scriptsize tp}$ denotes that this is the test-particle part of the linearized collision operator $C^l$ discussed in Section~\ref{sec:coll_op}}. Here (and throughout the rest of this paper), a bar denotes a quantity normalized to its reference value (i.e, $\bve=v_{\t{e}}/\rve$), $\rx=\ry/\gamma=v/\rve$ is the normalized speed, $c_C = 3\sqrt{\pi}\bnuee/4$, $c_\xi = \zeff+\Phi-\Psi+\bve^2\rdelta^4\rx^2/2$, $\zeff$ is the effective ion charge, $\Phi = \Phi(\rx/\bve)$ and $\Psi \!=\! \Psi(\rx/\bve)\!=\!\bve^2[\Phi-\bve^{-1}\rx\mbox{d}\Phi/\mbox{d}(\rx/\bve)]/2\rx^2$ are the error and Chandrasekhar functions, respectively, and $\rdelta\!=\!\rve/c$ (with $c$ the speed of light) is assumed to be a small parameter (i.e the thermal population is assumed to be non-relativistic). 

Changes to the plasma temperature manifest as shifts in the relative magnitude of the various terms in \Eq{eq:C} (through $\rdelta$ and the quantities with a bar), as well as a change in the overall magnitude of the operator, whereas changes in density only have the latter effect. In both cases, the distribution is effectively colliding with (and relaxing towards) a Maxwellian different from the one native to the reference momentum grid. Heat or particles are introduced to (or removed from) the bulk of the distribution when using this scheme, as all changes to plasma parameters are described by changes to the Maxwellian. This provides a powerful way of simulating rapid cooling, for instance associated with a tokamak disruption.

\subsection{Hot-tail runaway-electron generation}
If the time scale of the initial thermal quench in a disruption event is short enough -- comparable to the collision time -- the tail of the initial Maxwellian electron distribution will not have time to equilibrate as the plasma cools. The particles in this supra-thermal tail may constitute a powerful source of runaway electrons, should a sufficiently strong electric field develop before they have time to reconnect with the bulk electrons. This process is known as \emph{hot-tail generation}, and can be the dominant source of runaways under certain conditions \cite{Harvey,Helander_hot-tail}. It has previously been investigated analytically or using Monte-Carlo simulations \cite{Helander_hot-tail,Smith} or purpose-built finite-difference tools \cite{Smith,SmithVer}. Using \CODE{} to model a temperature drop enables the efficient study of a wider range of scenarios, and allows full use of other capabilities of \CODE{}, such as avalanche generation or synchrotron radiation reaction. Here, we restrict ourselves to a proof-of-principle demonstration, and leave a more extensive investigation to future work.

\begin{figure}
\begin{center}
	\includegraphics[width=0.39\textwidth, trim={0cm 0cm 0cm 0.5cm},clip]{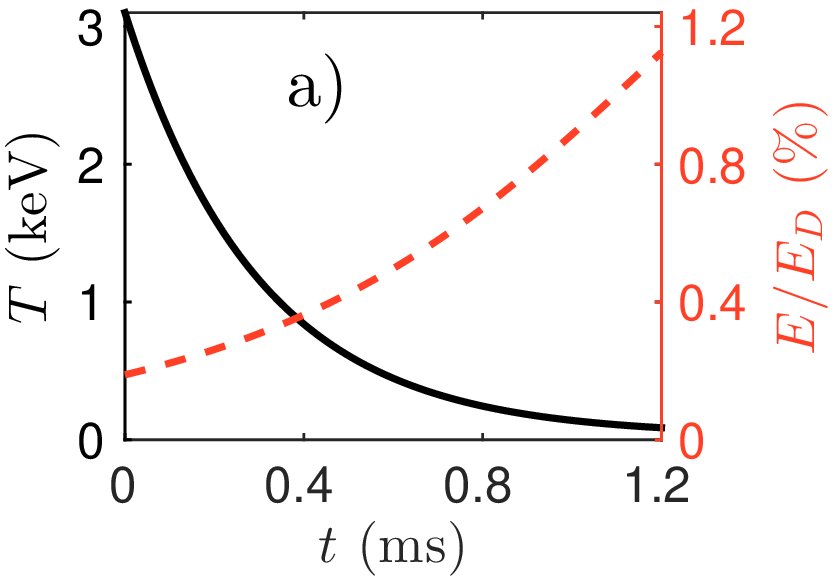}\hfill
	\includegraphics[width=0.59\textwidth, trim={0cm 0cm 1cm 0cm},clip]{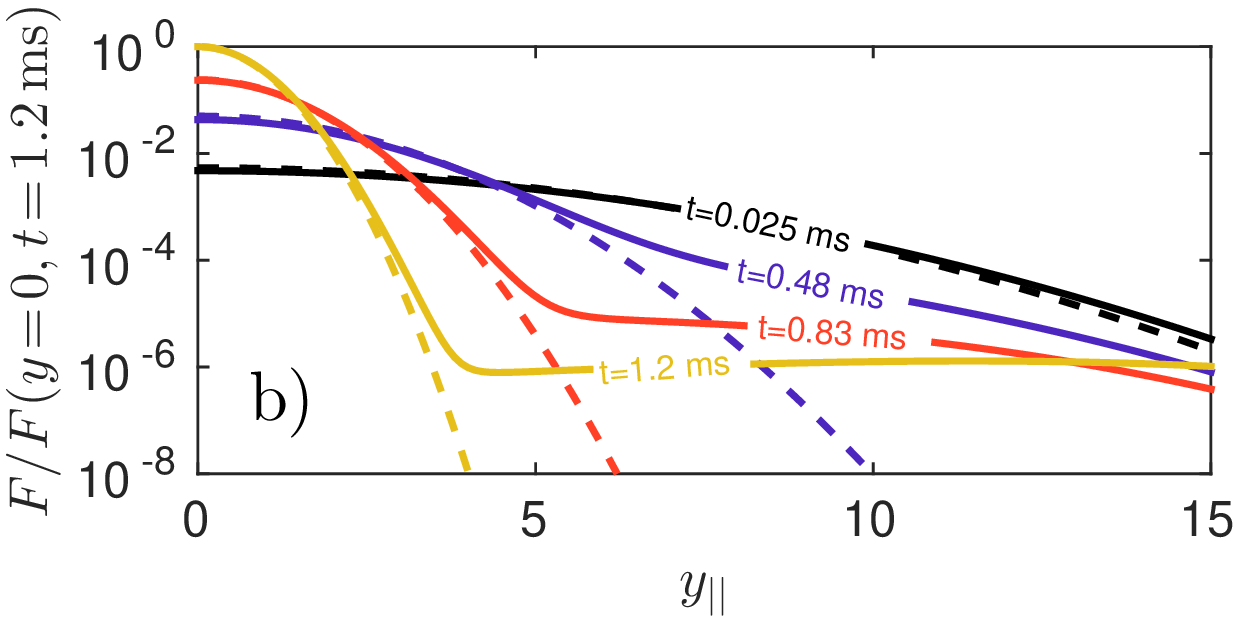}
	\caption{a) Temperature and electric-field evolution in Eqs.~(\ref{eq:T_evo}) and (\ref{eq:E_evo}). b) Parallel ($\xi\!=\!1$) electron distributions (solid) and corresponding Maxwellians (dashed) at several times during the temperature drop in a). A momentum grid with a fixed reference temperature $\reference{T}=100\,$eV was used and the distributions are normalized to $F(y\!=\!0)$ in the final time step to facilitate a comparison.  \label{fig:hot-tail_dist}}
\end{center}

\end{figure}

To facilitate a comparison to the theoretical work by Smith and Verwichte \cite{SmithVer}, we will model a rapid exponential temperature drop, described by
\begin{equation}
	T(t) = T_{\t{f}} + (T_0 - T_{\t{f}}) e^{-t/t_\star},
	\label{eq:T_evo}
\end{equation}
with $T_0=3.1\,$keV the initial temperature, $T_{\t{f}}=31\,$eV the final temperature, and $t_\star=0.3\,$ms the cooling time scale. We also include a time-dependent electric field described by 
\begin{equation}
	\frac{E(t)}{E_{\t{D}}} = \left(\frac{E}{E_{\t{D}}}\right)_{\!0}\sqrt{\frac{T_0}{T(t)}}, \label{eq:E_evo}
\end{equation}
with $(E/E_{\t{D}})_0 \!=\! 1/530$ the initial normalized electric field. The temperature and electric-field evolutions are shown in \Fig{fig:hot-tail_dist}a and are the same as those used in Fig.~5 of \Ref{SmithVer}, as are all other parameters in this section. 

Figure \ref{fig:hot-tail_dist}b, in which the additional parameters $n=2.8\power{19}\,$m$^{-3}$, and $\zeff=1$ were used, illustrates the distribution-function evolution during the temperature drop. The figure shows that as the temperature decreases, most of the electrons quickly adapt. At any given time $t$, the bulk of the distribution remains close to a Maxwellian corresponding to the current temperature $T(t)$. The initially slightly more energetic electrons, although part of the original bulk population, thermalize less efficiently. On the short cooling time-scale, they remain as a distinct tail, and as the thermal speed decreases they become progressively less collisional. This process is evident in the first three time steps shown ($t\!=\!0.025$--$0.83\,$ms). In the final time step, the electric field has become strong enough to start to affect the distribution, and a substantial part of the high-energy tail is now in the runaway region. This can be seen from the qualitative change in the tail of the distribution, which now shows a positive slope associated with a strong flow of particles to higher momenta.

For the temperature evolution in \Eq{eq:T_evo}, analytical results for the hot-tail runaway generation were obtained in \Ref{SmithVer}. Assuming the background density to be constant, the runaway fraction at time $t$ can be written as 
\begin{equation}
	\frac{n_{\t{r,dir}}}{n} = \frac{4}{\sqrt{\pi}} \int_{u_{\t{c}}}^\infty 
			\left[
				1 - \frac{(u_{\t{c}}^3-3\tau)^{2/3}}{(u^3-3\tau)^{2/3}}
			\right] 	
			e^{-u^2}u^2\d u,
			\label{eq:Smith18}
\end{equation}
where $\tau(t)\!=\!(3\sqrt{\pi}/4)\nuee(t-t_\star)\!=\!(3\sqrt{\pi}/4)(\tH-\tH_\star)$ is a normalized time, $u(t)\!=\!x^{[0]}+3\tau(t)$, $x^{[0]}$ is the speed normalized to the initial thermal speed, and $u_{\t{c}}$ is related to the critical speed for runaway generation: $u_{\t{c}}(t)\!=\!x^{[0]}_{\t{c}} +3\tau(t)$. Equation (\ref{eq:Smith18}), which corresponds to Eq.~(18) in \Ref{SmithVer}, is only valid when a significant temperature drop has already taken place (as manifested by the appearance of the cooling time scale $t_\star$ as a "delay" in the expression for $\tau$, see \cite{SmithVer}). Equation~(\ref{eq:Smith18}) is derived in the absence of an electric field; only an exponential drop in the bulk temperature is assumed. The electric field shown in \Fig{fig:hot-tail_dist}a is only used to define a runaway region, so that the runaway fraction can be calculated. In other words, it is assumed that the electric field does not have time to influence the distribution significantly during the temperature drop.

The runaway fraction calculated using \Eq{eq:Smith18} includes only the electrons in the \blue{actual} runaway region, i.e.~\blue{particles whose trajectories (neglecting collisional momentum-space diffusion) are not confined to a region close to the origin. In this case, the lower boundary of the runaway region is given in terms of the
limiting (non-relativistic) momentum $y$ for a given $\xi$: $y\!\geq\! y_{\t{c}\xi}\!=\!(\delta^2[(\xi+1)E/2E_{\t{c}}-1])^{-1/2}$ \cite{Smith}, where $E_{\t{c}}=4\pi e^3n\ln\Lambda/mc^2$ is the critical electric field for runaway generation \cite{ConnorHastie}}.
	The temperature drop does however lead to an isotropic high-energy tail (in the absence of an electric field). By \blue{defining the runaway region as $y>y_{\t{c}}=(\delta^2[E/E_{\t{c}} -1])^{-1/2}$, thereby} including all particles with $v>v_{\t{c}}$, \Eq{eq:Smith18} can be simplified to 
\begin{equation}
	\frac{n_{\t{r}}}{n} = \frac{2}{\sqrt{\pi}}u_{\t{c}} e^{-u_{\t{c}}^2} + \mbox{erfc}(u_{\t{c}}),
	\label{eq:Smith19}
\end{equation}
where erfc$(x)$ is the complementary error function. \blue{By default, \CODE{} uses such an isotropic runaway region, which is a good approximation in the case of only Dreicer and avalanche generation (especially once the runaway tail has become substantial); however, in the early stages of hot-tail-dominated scenarios, the isotropic runaway region significantly overestimates the actual runaway fraction, and the lower boundary $y_{\t{c}\xi}$ must be used.}

\begin{figure}
\begin{center}
\includegraphics[width=1\textwidth, trim={0cm 0cm 0.1cm 0.45cm},clip]{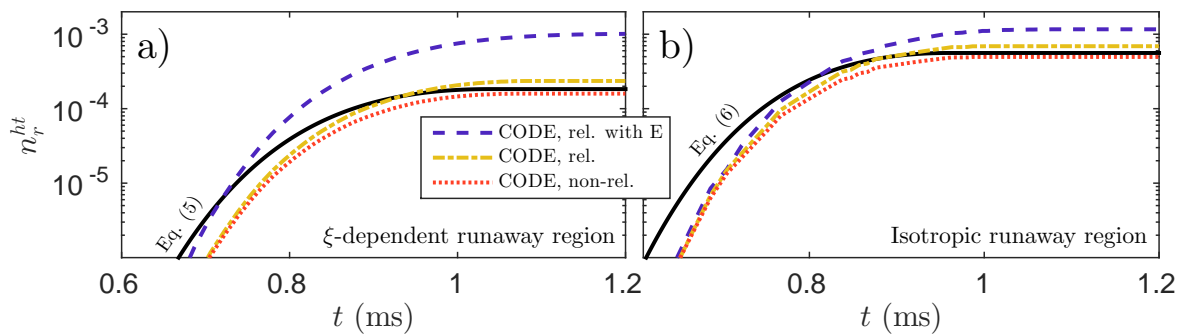}
\caption{Hot-tail runaway density obtained using \CODE{} -- with \blue{(blue, dashed)} and without \blue{(yellow, dash-dotted; red, dotted)} an electric field included during the temperature drop -- and the analytical estimates Eqs. (\ref{eq:Smith18}) and (\ref{eq:Smith19}) \blue{(black, solid)}, for the temperature and E-field evolution in Fig.~\ref{fig:hot-tail_dist}a. \blue{An a) $\xi$-dependent and b) isotropic lower boundary of the runaway region was used}. The collision operator in Eq.~(\ref{eq:C}) was used for the \blue{blue and yellow lines}, whereas its non-relativistic limit was used for the \blue{red and black lines}.}
\label{fig:hot-tail}
\end{center}
\end{figure}

Figure \ref{fig:hot-tail} compares the runaway density evolution computed with \CODE{}\blue{, using both $\xi$-dependent and isotropic runaway regions,} to Eqs.~(\ref{eq:Smith18}) and (\ref{eq:Smith19})\blue{, respectively}. \blue{The parameters of the hot-tail scenario shown in \Fig{fig:hot-tail_dist} were used, and} no avalanche source was included in the calculation. The collision operator used in \Ref{SmithVer} is the non-relativistic limit of \Eq{eq:C}, with $c_\xi\!=\!0$ (since the distribution is isotropic in the absence of an electric field). \CODE{} results using both this operator (red, \blue{dotted}) and the full \Eq{eq:C} (yellow, \blue{dash-dotted}) are plotted in \Fig{fig:hot-tail}, with the latter producing $\!~\!\sim 50\%$ more runaways in total. This difference can likely be explained by the relatively high initial temperature (3~keV) in the scenario considered, in which case the non-relativistic operator is not strictly valid for the highest-energy particles. Good agreement between \CODE{} results and \blue{Eqs.~(\ref{eq:Smith18}) and (\ref{eq:Smith19}) (black, solid)} is seen for the saturated values in the figure. A \CODE{} calculation where the electric-field evolution is properly included in the kinetic equation (corresponding to the distribution evolution in \Fig{fig:hot-tail_dist}b) is also \blue{included (blue, dashed)}, showing increased runaway production. \blue{With the isotropic runaway region (\Fig{fig:hot-tail}b), the increase is smaller than a factor of 2, and} neglecting the influence of the electric field can thus be considered reasonable \blue{for the parameters used}, at least for the purpose of gaining qualitative understanding. \blue{With the $\xi$-dependent runaway region (Fig.~\ref{fig:hot-tail}a), the change in runaway generation is more pronounced, and the inclusion of the electric field leads to an increase by almost an order of magnitude. Note that the final runaway density with the electric field included is very similar in Figs.~\ref{fig:hot-tail}a and \ref{fig:hot-tail}b, indicating that the details of the lower boundary of the runaway region become unimportant once the tail is sufficiently large. Throughout the remainder of this paper we will make use of the isotropic runaway region.} 

\blue{We conclude that, in order to} obtain quantitatively accurate results, the electric field should be properly included, and a relativistic collision operator should be used. This is especially true when modelling ITER scenarios, where the initial temperature \blue{can be} significantly higher than the 3 keV used here.

\section{Conservative linearized Fokker-Planck collision operator}
\label{sec:coll_op}
Treating the runaway electrons as a small perturbation to a Maxwellian distribution function, the Fokker-Planck operator for electron-electron collisions \cite{Landau,RosMacJudd} can be linearized and written as
$C\{f\} \!\simeq\! C^l\{f\} \!=\! \Ctp  \!+\! \Cfp$. The so-called \emph{test-particle term}, $\Ctp = \Cnl\{f_1,f_{\t{M}}\}$, describes the perturbation colliding with the bulk of the plasma, whereas the \emph{field-particle term}, $\Cfp=\Cnl\{f_{\t{M}},f_1\}$, describes the reaction of the bulk to the perturbation. Here $\Cnl$ is the non-linear Fokker-Planck-Landau operator, $f_{\t{M}}$ denotes a Maxwellian, and $f_1\!=\!f\!-\!f_{\t{M}}$ the perturbation to it ($f_1\! \ll\! f_{\t{M}}$). Collisions described by $C\{f_1,f_1\}$ are neglected since they are second order in $f_1$. The full linearized operator $C^l$ conserves particles, momentum and energy. Since it is proportional to a factor $\exp(-y^2)$, the field-particle term mainly affects the bulk of the plasma, and is therefore commonly neglected when studying runaway-electron kinetics. The test-particle term in \Eq{eq:C} only ensures the conservation of particles, however, not momentum or energy.

Under certain circumstances, it is necessary to use a fully conservative treatment also for the runaway problem, in particular when considering processes where the conductivity of the plasma is important. In the study of runaway dynamics during a tokamak disruption using a self-consistent treatment of the electrical field, accurate plasma-current evolution is essential, and the full linearized collision operator must be used. A non-linear collision operator valid for arbitrary particle (and bulk) energy has been formulated \cite{BelBud,BraamsKarney}. The collision operator originally implemented in \CODE{} is the result of an asymptotic matching between the highly relativistic limit of the test-particle term of the linearized version of that operator, with the usual non-relativistic test-particle operator \cite{Papp2011}, and is given in Eq.~(\ref{eq:C}). The relativistic field-particle term is significantly more complicated, however, and its use would be computationally more expensive. Here we instead implemented the non-relativistic field-particle term, as formulated in Refs.~\cite{CattoTsang,LiErnst}. As will be shown, this operator (together with the non-relativistic limit of Eq.~\ref{eq:C}) accurately reproduces the Spitzer conductivity for sufficiently weak electric fields and temperatures where the bulk is non-relativistic. Using the normalization in Section~\ref{sec:td}, the field-particle term is
\begin{equation}
\hat{C}^{\mbox{\scriptsize fp}} 
= \frac{c_C}{\pi^{3/2}} e^{-\bve^{-2}\rx^2}
	\left[ 
		\frac{2\rx^2}{\bve^4} \frac{\partial^2 G}{\partial \rx^2} - \frac{2}{\bve^2}H + 4\pi F
	\right],
	\label{eq:fp}
\end{equation}
where $G$ and $H$ are the Rosenbluth potentials, obtained from the distribution using
\begin{equation}
\rve^{2} \nabla_{\!\bf v}^2 H = -4\pi F, \qquad 
\rve^{2} \nabla_{\!\bf v}^2 G = 2 H. 
\label{eq:Poisson}
\end{equation}
The system of equations composed of Eqs.~(\ref{eq:fp}-\ref{eq:Poisson}), together with the non-relativistic limits of Eqs.~(\ref{eq:kinetic}-\ref{eq:C}) ($\ry\!\to\!\rx$ and $\rdelta\!\to\! 0$), is discretized (see Ref.\cite{CODE}) and solved using an efficient method described in \Ref{LandremanErnst}. The equations are combined into one linear system of the form
\begin{equation}
	\left(\begin{array}{ccc}
		M_{11} & M_{12} & M_{13} \\
		M_{21} & M_{22} & 0 \\
		0      & M_{31} & M_{33}
	\end{array}\right)
	\left(\begin{array}{c}
		F \\
		G \\
		H 
	\end{array}\right)
	=
	\left(\begin{array}{c}
		S_i \\
		0 \\
		0 
	\end{array}\right),	
\end{equation}
where the first row describes the kinetic equation (\ref{eq:kinetic}) (with $S_i$ representing any sinks or sources), and the second and third rows correspond to \Eq{eq:Poisson}. This approach makes it possible to consistently solve for both the Rosenbluth potentials and the distribution with a single matrix operation. Since there is no explicit need for the Rosenbluth potentials, however, $G$ and $H$ can be eliminated by solving the block system analytically:
\begin{equation}
	\left( 
		M_{11} - \left[M_{12} - M_{13}M_{33}^{-1}M_{32}\right]M_{22}^{-1}M_{21}
	\right)F \equiv MF = S_i.
\end{equation}  
If only the test-particle operator (Eq.~\ref{eq:C}) is used, $M$ reduces to $M_{11}$.
Since the Rosenbluth potentials are defined through integrals of the distribution, the field-particle term introduces a full block for each Legendre mode into the normally sparse matrix describing the system.  However, the integral dependence on $F$ also implies that significantly fewer modes are required to accurately describe the potentials (compared to $F$), and the additional computational cost is modest (the operator $\nabla_{\!\bf v}^2$ is proportional to $l^2$, with $l$ the Legendre mode index, and $G$ and $H$ therefore decay rapidly with increasing $l$).

\begin{figure}
\includegraphics[width=1\textwidth]{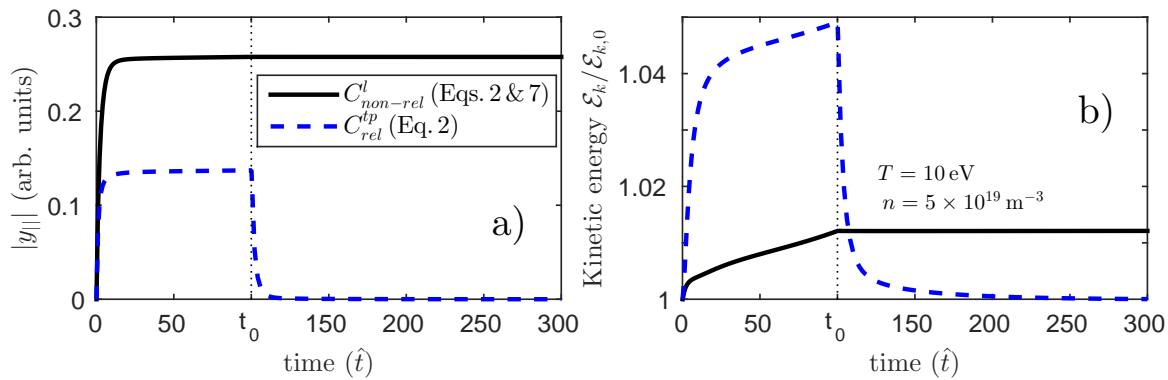}
\caption{a) Parallel momentum and b) energy moments of the distribution function in \CODE{}, using different collision operators.  Initially, $E\!=\!50$~V/m and $\zeff\!=\!1$ were used, but for $t\!>\!t_0$, the electric field was turned off and the ion charge set to $\zeff\!=\!0$. 
Using two Legendre modes for the field-particle term was sufficient to achieve good conservation of energy and parallel momentum.}
\label{fig:conservation}
\end{figure}

The conservation properties of the full non-relativistic collision operator (Eqs.~\ref{eq:C} and \ref{eq:fp}), as well as the relativistic test-particle operator in \Eq{eq:C}, are shown in \Fig{fig:conservation}. As an electric field is applied to supply some momentum and energy to the distribution, the parallel momentum (\Fig{fig:conservation}a) quickly reaches a steady-state value corresponding to the plasma conductivity, which differs by about a factor of two for the two operators (see below). The electric field is turned off at $t\!=\!t_0\!=\!100$ collision times (and $\zeff\!=\!0$ is imposed to isolate the behavior of the electron-electron collision operator), at which point the parallel momentum for the operator in \Eq{eq:C} (blue, dashed) is lost on a short time scale as the distribution relaxes back towards a Maxwellian. In contrast, the full linearized operator (black, solid) conserves parallel momentum in a pure electron plasma, as expected. 

The electric field continuously does work on the distribution, a large part of which heats the bulk electron population, but the linearization of the collision operator breaks down if the distribution deviates too far from the equilibrium solution. As long as a non-vanishing electric field is used together with an energy conserving collision operator, an adaptive sink term removing excess heat from the bulk of the distribution must be included in \Eq{eq:kinetic} to guarantee a stable solution. Physically this accounts for loss processes that are not included in the model, such as line radiation, bremsstrahlung and radial heat transport. The magnitude of the black line in \Fig{fig:conservation}b therefore reflects the energy content of the runaway population -- not the total energy supplied by the electric field -- since a constant bulk energy is enforced. The energy sink is not included for $t\!>\!t_0$ (since $E\!=\!0$), however, and the energy conservation observed is due to the properties of the collision operator itself. Again, the use of the collision operator in \Eq{eq:C} is associated with a quick loss of kinetic energy as soon as the electric field is removed.

\begin{figure}
\includegraphics[width=0.59\textwidth, trim={0cm 0cm 1cm 0cm},clip]{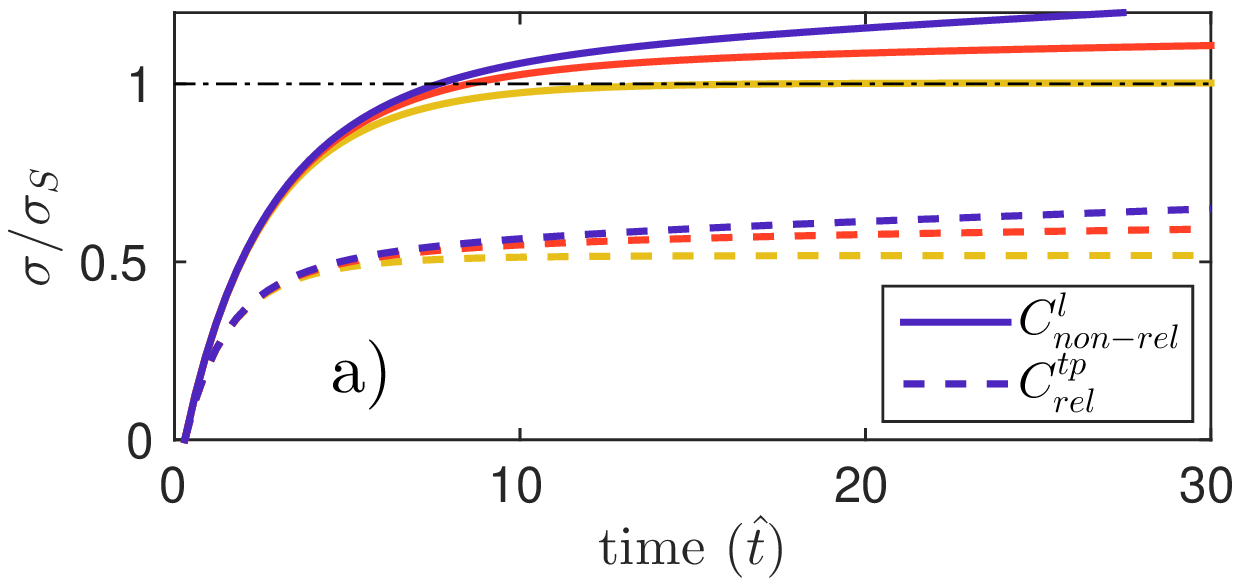}\hfill
\includegraphics[width=0.39\textwidth, trim={0cm 0cm 0cm 0cm},clip]{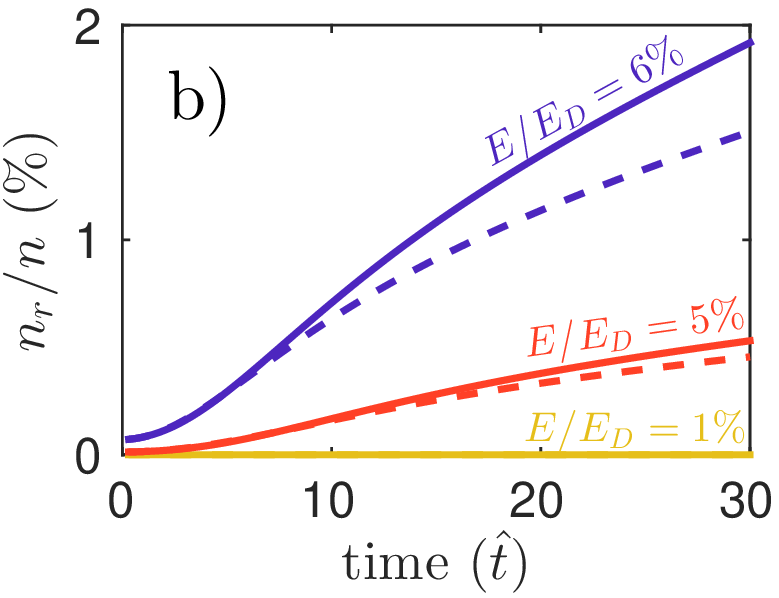}
\caption{a) Conductivity (normalized to the Spitzer value) and b) normalized runaway density, as functions of time for different collision operators (non-relativistic full linearized: solid; relativistic test-particle: dashed) and E-field strengths ($E/E_{\t{D}}\!=\!1$\%: yellow; $E/E_{\t{D}}\!=\!5$\%: red; $E/E_{\t{D}}\!=\!6$\%: blue), considering only Dreicer runaway generation. The parameters $T\! =\! 1$~keV, $n\!=\!5\power{19}\,$m$^{-3}$ and $\zeff\!=\!1$ were used.}
\label{fig:Spitzer}
\end{figure}

The electrical conductivity of a fully ionized plasma subject to an electric field well below the Dreicer value -- the \emph{Spitzer conductivity} -- can be expressed as 
\begin{equation}
	\sigma_{\t{S}} = L(\zeff) \frac{ne^2}{\zeff\, m \nuee},
\end{equation}
where $L(\zeff)$ is a transport coefficient which takes the value $L\!\simeq\! 2$ in a pure hydrogen plasma \cite{SpitzerHarm}. Figure~\ref{fig:Spitzer} demonstrates that the conductivity calculated with \CODE{} reproduces the Spitzer value for moderate electric-field strengths, if the conservative collision operator is used, and the initial Maxwellian adapts to the applied electric field on a time scale of roughly 10 collision times. For field strengths significantly larger than $E_{\t{c}}$, the conductivity starts to deviate from $\sigma_S$, as a runaway tail begins to form (\Fig{fig:Spitzer}b); in this regime, the calculation in \Ref{SpitzerHarm} is no longer valid. Using the collision operator in \Eq{eq:C} consistently leads to a conductivity which is lower by about a factor of 2, as expected (see for instance \Ref{HelanderSigmar}). The runaway growth is also affected, with the conserving operator leading to a larger runaway growth rate.

\section{Improved operator for knock-on collision}
\label{sec:knock-on}

The Fokker-Planck collision operators discussed in Section~\ref{sec:coll_op} accurately describe grazing collisions -- small-angle deflections which make up the absolute majority of particle interactions in the plasmas under consideration. Large-angle collisions are usually neglected as their cross section is significantly smaller, but in the presence of runaway electrons they can play an important role in the momentum space dynamics, as an existing runaway can transfer enough momentum to a thermal electron in one collision to render it a runaway, while still remaining in the runaway region itself. Such \emph{knock-on} collisions can therefore lead to an exponential growth of the runaway density -- an \emph{avalanche} \cite{RosPut,Jayakumar}.

\begin{figure}
\includegraphics[height=0.27\textheight, trim={0cm 0.25cm 1.8cm 0.2cm},clip]{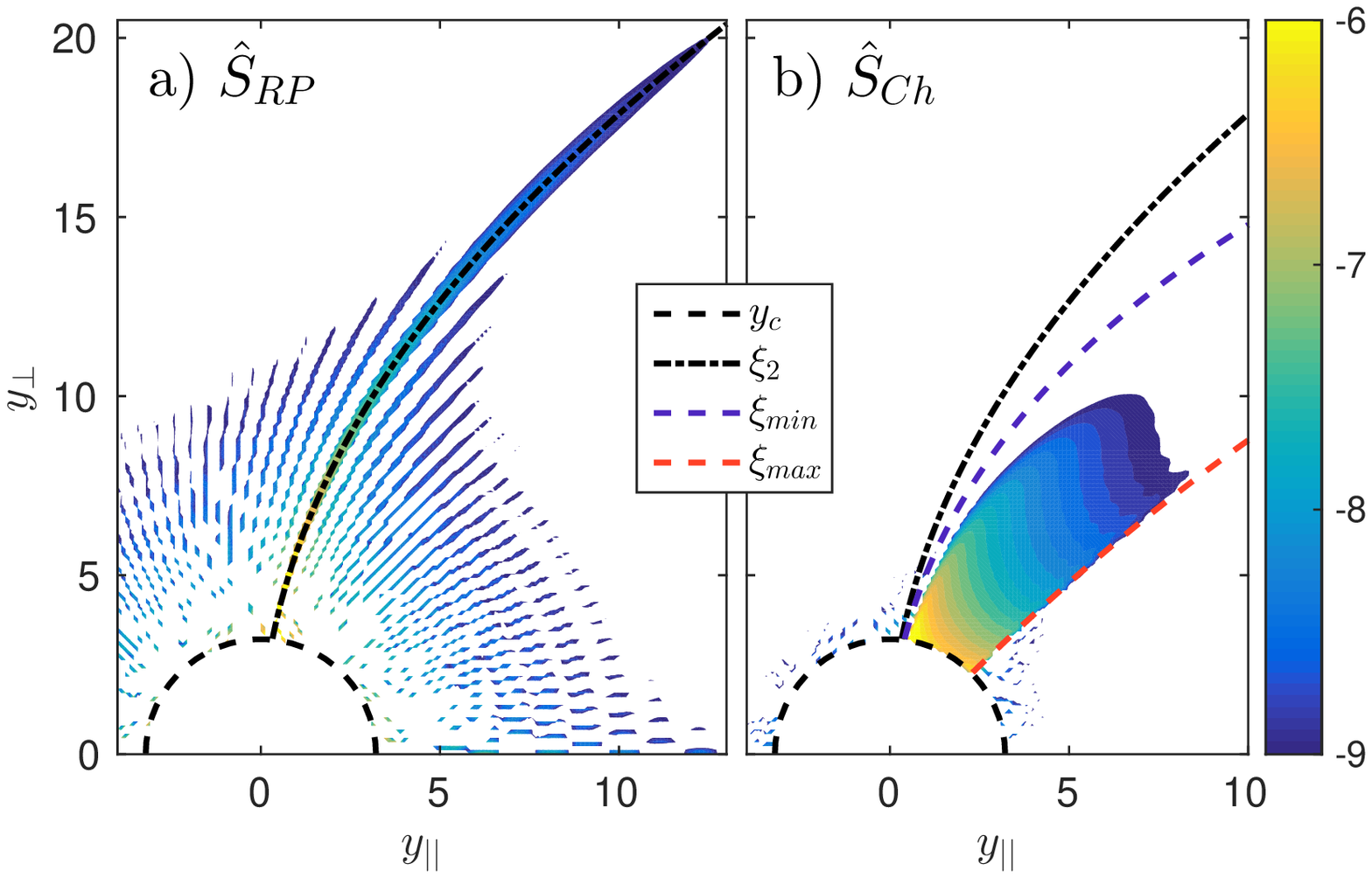}\hfill
\includegraphics[height=0.27\textheight, trim={0cm 0.1cm 0.7cm 0.5cm},clip]{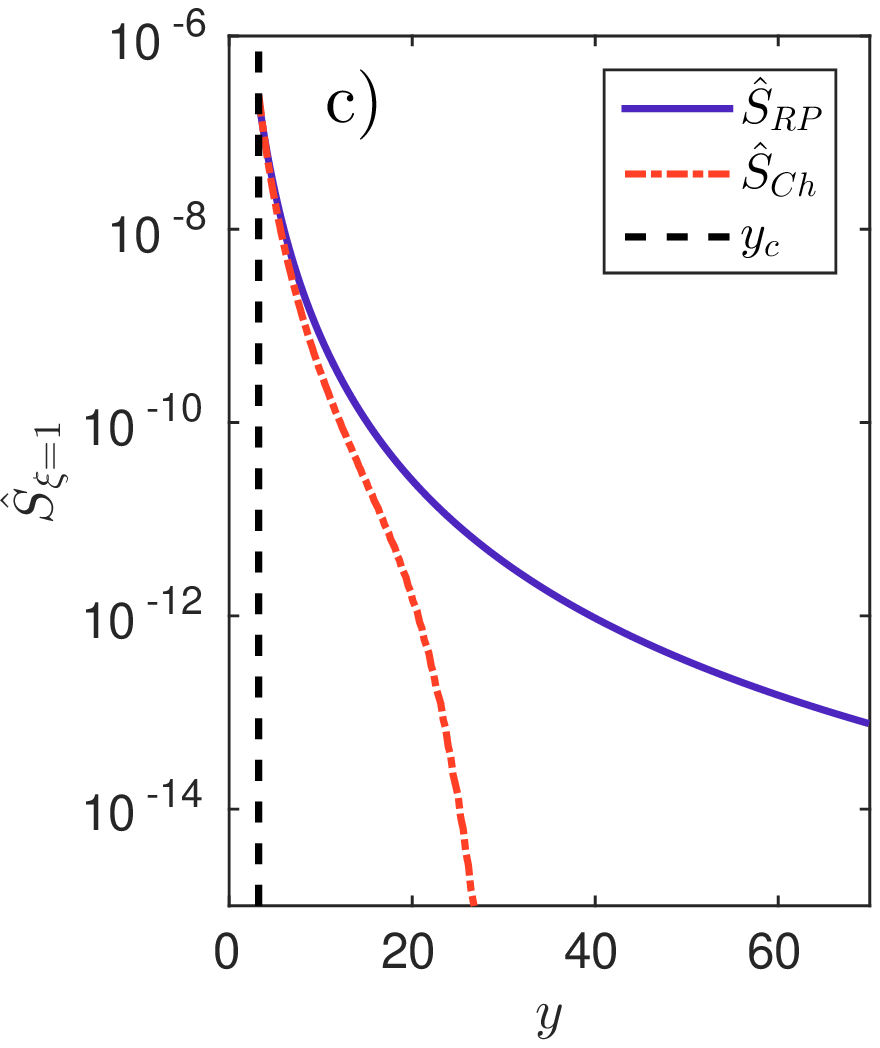}
\caption{Contour plots of the magnitude of the source in a) \Eq{eq:RP} and b) \Eq{eq:Ch} in ($y_\|$,$y_\perp$) momentum space, given the same electron distribution. The plotted quantity is $\log_{10}{\hat{S}}$ and $y_{\t{c}}$ defines the lower bound of the runaway region. The angle-averaged source magnitudes are shown in c). The parameters $T = 1$~keV, $n=5\cdot 10^{19}$~m$^{-3}$, $\zeff=1$ and $E=1$~V/m, with $\max(y)=70$, were used to obtain the distribution, and the simulation was run for 300 collision times with primary generation only. }
\label{fig:sources}
\end{figure}

In the absence of a complete solution to the Boltzmann equation, we model avalanche runaway generation using an additional source term in the kinetic equation (\ref{eq:kinetic}), evaluated for $\ry\!>\!\ry_{\t{c}}$. A commonly used operator was derived by Rosenbluth and Putvinski \cite{RosPut} and takes the form
\begin{equation}
\hat{S}_{\RP} = \frac{n_{\t{r}}}{n}\, \normalized{n}^2 
\left[
	\frac{3\pi\rdelta^3}{16\rlnL}\delta_{\t{D}}(\xi-\xi_{2})\frac{1}{\ry^{2}}\pder{}{\ry}\left(\frac{1}{1-\sqrt{1+\rdelta^{2}\ry^{2}}}\right)
\right],
\label{eq:RP}
\end{equation}
where $n_{\t{r}}$ is the number density of runaway electrons, $\bar{n}$ is the density normalized to its reference value, and $\delta_{\t{D}}$ is the Dirac $\delta$-function. In the derivation, the momentum of the incoming particle is assumed to be very large (simplifying the scattering cross section) and its pitch-angle vanishing ($\xi\!=\!1$). It is also assumed that the incoming particle is unaffected by the interaction.  These conditions imply that the generated \emph{secondary} particles are all created on the curve $\xi\!=\!\xi_2\!=\!\rdelta\ry/(1+\sqrt{1+\rdelta^2\ry^2})$ (which is a parabola in $[y_\|,y_\perp]$-space), and that \emph{all} runaways (from the point of view of the avalanche source) are assumed to have momentum $p\!=\!\g v/c\!=\!\delta y\gg 1$ (since $\hat{S}_{\RP}\!\propto\! n_{\t{r}}$). They can therefore contribute equally strongly to the avalanche process. This has the peculiar and non-physical consequence that particles can be created with an energy higher than that of any of the existing runaways. The $\delta$-function in $\xi$ is numerically ill-behaved, as it produces significant oscillations (Gibbs phenomenon) when discretized using the Legendre-mode decomposition employed in \CODE{} (see Fig.~\ref{fig:sources}a).

An operator that relaxes the assumption of very large runaway momentum has been presented by Chiu et al. \cite{Chiu}. It has the form
\begin{equation}
\hat{S}_{\Ch}(\ry,\xi) = \normalized{n}\, \frac{2\pi e^4}{m^2c^3}\,
	\frac{\rn \rdelta^{3}}{\rnuee}\,
	\frac{\rx}{\ry^{2}\xi}
	\left(\ry_{\tin}\right)^{4} F^\star(\ry_{\tin})\,\Sigma\left(\g,\gin\right),
	\label{eq:Ch}
\end{equation}
where \pagebreak
\begin{eqnarray}
	\Sigma(\g,\gin) &=& \frac{ \gin^2 }{ (\gin^2-1)(\g-1)^2(\gin-\g)^2 }
	\left[
		(\gin-1)^2 \right.\nonumber \\
		&\qquad &- \left.\frac{(\g-1)(\gin-\g)}{\gin^2}
		\left(
			2\gin^2 + 2\gin -1 -(\g-1)(\gin-\g)
		\right)
	\right] \label{eq:Moller}
\end{eqnarray} 
is the M\o ller scattering cross section \cite{Moller} and $F^\star$ is the pitch-angle-averaged distribution of incoming runaways with properties $\ry_{\tin}$ and $\gamma_{\tin}$. All incoming particles are thus still assumed to have zero pitch angle ($\xi\!=\!1$), but their energy distribution is properly taken into account. In \CODE{}, $F^\star$  is computed from the 0th Legendre mode of $F$; $F^\star\!=\!2F_{0}$.

From the conservation of 4-momentum in a collision, the momentum-space coordinates are related through 
\begin{equation}
\xi = \sqrt{ \frac{ (\gamma-1)(\gamma_{\tin}+1) }{ (\gamma+1)(\gamma_{\tin}-1) } }, 
\end{equation}
which restricts the region where the source is non-vanishing (this relation is analogous to the parabola $\xi_2$ in the case of the operator in Eq.~\ref{eq:RP}). Since the electrons participating in a collision are indistinguishable, it is sufficient to consider only the cases where the energy of the created secondary runaway is less than half of the primary energy, $(\gamma\!-\!1)\!\leq\!(\gamma_{\tin}\!-\!1)/2$, which with the above equation leads to the condition $\xi\leq\xi_{\max}=\sqrt{\gamma/(\gamma+1)}$. By the same argument, the maximum attainable runaway energy in the simulation (the maximum of the momentum grid) leads to the condition $\xi\geq\xi_{\min} = \sqrt{(\gamma-1)(\gamma_{\tmax}+1)/(\gamma+1)(\gamma_{\tmax}-1)}$.

The magnitudes of the two sources (\ref{eq:RP}) and (\ref{eq:Ch}) are computed from a given typical runaway distribution function, and shown in \Fig{fig:sources}a and b. Curves corresponding to the \blue{parabola} $\xi_2$, as well as the limits $\xi_{\min}$ and $\xi_{\max}$ are also included. Note that the amount of numerical noise is significantly reduced for the source in \Eq{eq:Ch}. In order to avoid double-counting the small-angle collisions described by the Fokker-Planck-Landau collision operator $C$, the knock-on source must be cut off at some value of momentum sufficiently far from the thermal bulk. As can be seen from the figure, however, the magnitude of both sources increases with decreasing momenta, and the avalanche growth rate is therefore sensitive to the specific choice of momentum cut-off. Since our particular interest is the generation of runaway electrons, we choose to place the cut-off at $y\!=\!y_{\t{c}}$, so that the sources are non-vanishing only in the runaway region \cite{CODE,Harvey}. Secondary particles deposited just below the threshold -- although not technically runaways -- could eventually diffuse into the runaway region, thereby potentially increasing the Dreicer growth rate. In Ref.~\cite{Nilsson}, such effects were however shown to be negligible for the operator in \Eq{eq:RP}, indicating that the vast majority of particles deposited at $y\! < \! y_{\t{c}}$ are slowed down rather than accelerated (as expected). This reduces the sensitivity of the avalanche growth rate to the choice of momentum cut-off (as long as $y_{\t{cut-off}}\!\leq\!y_{\t{c}}$), and reaffirms our choice $y_{\t{cut-off}}\!=\!y_{\t{c}}$.

 \Fig{fig:sources}c shows the source terms integrated over pitch-angle, and as expected, the source in \Eq{eq:Ch} extends only up to $y \simeq y_{\max}/2$, whereas the source in \Eq{eq:RP} is non-vanishing also for larger momenta. The amount of secondary runaways generated by the two sources agrees well at low energies, but less so further away from the bulk. In this particular case, the total source magnitude $\int\! \hat{S}\, y^2\mbox{d}y\mbox{d}\xi$ agrees to within 25\%, as most of the secondaries are created close to the boundary of the runaway region. 

\subsection{Avalanche growth rates for the different operators}
In general, the avalanche growth rate produced by the two sources can differ substantially. We will illustrate this point by considering the M\o ller cross section in more detail. We choose to quantify the source magnitude for an arbitrary distribution by computing the cross section, integrated over the energy of the outgoing (secondary) particle and normalized to $r_0^2$, with $r_0$ the classical electron radius. In other words, we look at the total normalized cross section for an incoming particle with $\gin$ to participate in a knock-on collision resulting in avalanche \cite{Aleynikov}: 
\begin{eqnarray}
	K_{\Ch}(\gin) &= \int_{\g_{\t{c}}}^{(\gin-1)/2+1} &\Sigma(\g,\gin)\mbox{d}\g \nonumber\\
			&= (\gin^2-1)^{-1}
				&\left[
					\frac{\gin^2}{\g_{\t{c}}-1} + \frac{\gin^2}{\g_{\t{c}}-\gin} \right. \nonumber\\
					&&\ + \left. \frac{2\gin-1}{\gin-1}\ln\left( \frac{\g_{\t{c}}-1}{\gin-\g_{\t{c}}} \right)
					+ \frac{\gin+1}{2} - \g_{\t{c}}
				\right],
\end{eqnarray}
where $\g_{\t{c}} = \sqrt{(E/E_{\t{c}})/(E/E_{\t{c}}-1)}$ corresponds to the critical momentum for runaway generation and the upper integration boundary stems from the condition leading to $\xi_{\max}$.
This expression is relevant to the source in \Eq{eq:Ch}, which uses the complete cross section (\ref{eq:Moller}), whereas for the more simple source in \Eq{eq:RP}, only the leading-order term in $\gin$ in the scattering cross section is taken into account. This corresponds to taking the high-energy limit of the above equation, so that
\begin{equation}
	K_{\RP} = \frac{1}{\g_{\t{c}} -1}
\end{equation}
becomes a simple constant.

\begin{figure}
\includegraphics[width=0.5\textwidth, trim={0cm 0cm 0.95cm 0.2cm},clip]{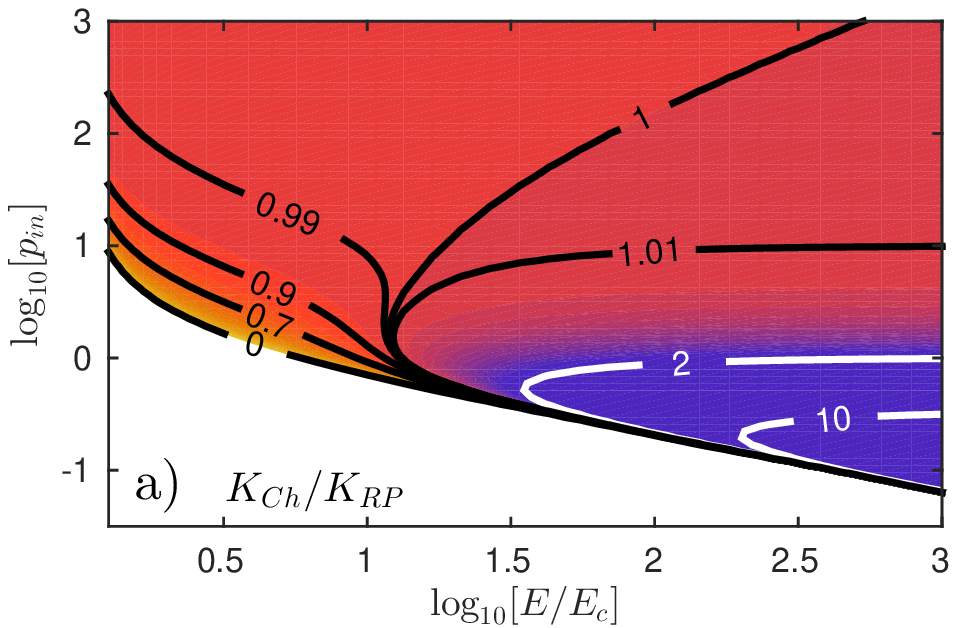}
\includegraphics[width=0.5\textwidth, trim={0cm 0cm 0.9cm 0.2cm},clip]{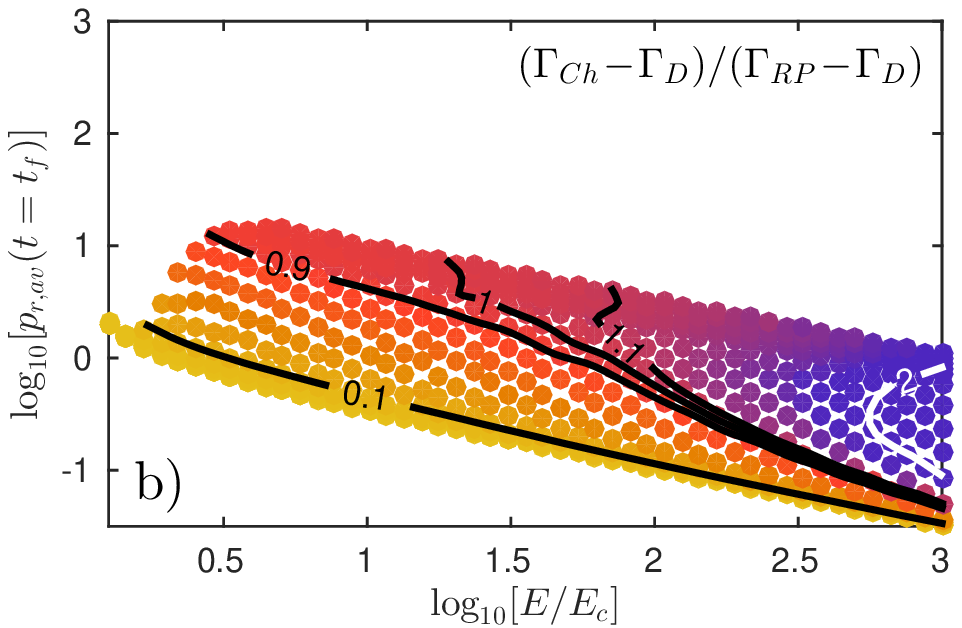}
\caption{a) Contours (black, white) of the ratio of total cross-sections ($K_{\Ch}/K_{\RP}$) for an electron with $p_{\tin}$ to contribute to the avalanche process, as a function of $p_{\tin}\!=\!\gin v_{\tin}/c\! =\! \sqrt{\gin^2-1}$ and $E/E_{\t{c}}$.
b) Ratio of avalanche growth rates ($[\Gamma_{\Ch}\!-\!\Gamma_{\Dr}]/[\Gamma_{\RP}\!-\!\Gamma_{\Dr}]$) in \CODE{} simulations. The parameters $T\!\in\! [0.1\,$eV$,5\,$keV$]$, $E/E_{\t{c}}\!\in\! [1.1,1000]$, $n\!=\!5\power{19}\,$m$^{-3}$ and $\zeff\!=\!1$ were used.}
\label{fig:K_ratio}
\end{figure}

To systematically explore the relative magnitude of the two sources, the ratio $K_{\Ch}/K_{\RP}$ is plotted in \Fig{fig:K_ratio}a. As expected, the two expressions agree very well at high primary momenta. At somewhat lower momenta, of the order $\g\!\approx\! p\! \lesssim\! 5$, two distinct regions are discernible. For $E/E_{\t{c}}\! \lesssim\! 10$ (the orange region), the simplified cross section is larger than the full expression, and the Rosenbluth-Putvinski operator (\ref{eq:RP}) is likely to overestimate the avalanche generation. For $E/E_{\t{c}}\! \gtrsim\! 10$, the opposite is true, and the operator in \Eq{eq:Ch} has a significantly larger cross section for $E/E_{\t{c}}\! \gtrsim\! 30$ (the blue region). The more accurate operator (\ref{eq:Ch}) should thus be expected to produce more runaways when the runaway population is at predominantly low energies, and $E/E_{\t{c}}$ is large. For both of these conditions to be fulfilled simultaneously (and at the same time avoid a slide-away scenario), the temperature must be low so that $E/E_{\t{D}}\ll 1$ even for large $E/E_{\t{c}}$. The effect is also likely to be most apparent at relatively early times, before the runaway tail has extended to multi-MeV energies. 

\CODE{} simulations support the above conclusions and show excellent qualitative agreement, as shown in \Fig{fig:K_ratio}b. The figure shows the ratio of final avalanche growth rates $(\Gamma_{\Ch}-\Gamma_{\Dr})/(\Gamma_{\RP}-\Gamma_{\Dr})$, with $\Gamma_i \!=\! n_{\t{r}}^{-1}(\d n_{\t{r}}/\d\tH)$ the growth rate obtained in a \CODE{} run using source $i$ (here the subscript D denotes pure Dreicer generation). Each marker in the figure is thus computed from three separate \CODE{} runs. As a proxy for $p_{\tin}$, the average runaway momentum $p_{\t{r,av}}$ in the final time step $t_{\t{f}}$ of the simulation without a source was used\blue{, and for a given $E/E_{\t{c}}$, different $p_{\t{r,av}}$ were obtained from simulations with varying values of $T$ (and corresponding values of $E/E_{\t{D}}$)}. The simulations were run for $t_{\max}\!=\!5000$ collision times, and $t_{\t{f}}$ was set to either $t_{\max}$, the first time step for which $n_{\t{r}}\!>\!5\%$, or the first time step in which the growth rate started to become affected by the proximity of the runaway tail to the end of the simulation grid, whichever occurred first. The parameters of the scan were chosen to focus on the most interesting region of \Fig{fig:K_ratio}a -- by performing longer simulations on larger momentum grids, the upper part of the figure could also be studied. Exact agreement between Figs. \ref{fig:K_ratio}a and b can not be expected, since the source, in addition to the cross section, depends on the details of the runaway distribution. Figure \ref{fig:K_ratio}a should thus be viewed as a simplified analytical estimate for \Fig{fig:K_ratio}b. The different regions identified in \Fig{fig:K_ratio}a are still apparent in \Fig{fig:K_ratio}b, however they are somewhat shifted in parameter space. In particular, the region where the Rosenbluth-Putvinski operator produces a higher growth rate is larger, whereas the opposite region -- where the operator in \Eq{eq:Ch} dominates -- is smaller, or at least shifted to higher values of $E/E_{\t{c}}$.

\begin{figure}
\begin{center}
	\includegraphics[width=0.5\textwidth, trim={0cm 0cm 0.9cm 0.2cm},clip]{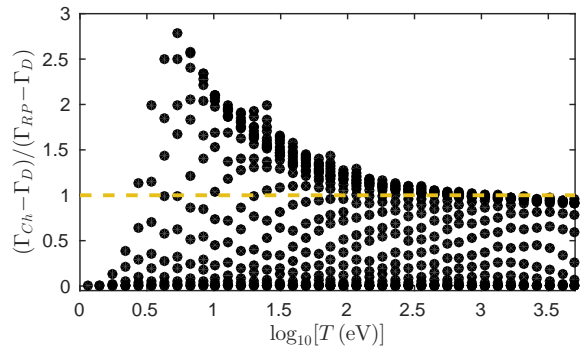}
\end{center}
\caption{Ratio of avalanche growth rates ($[\Gamma_{\Ch}\!-\!\Gamma_{\Dr}]/[\Gamma_{\RP}\!-\!\Gamma_{\Dr}]$) in \CODE{} simulations, as a function of temperature. The same parameters as in \Fig{fig:K_ratio} were used.}
\label{fig:Growth_vs_temp}
\end{figure}

Figure \ref{fig:Growth_vs_temp} shows all the data points in \Fig{fig:K_ratio}b, as a function of temperature. The figure confirms that the region where the more accurate operator produces a significantly higher growth rate is only accessible at temperatures $T < 100\,$eV (in the domain of validity of a linearized treatment). As is evident in the figure, however, regions where the Rosenbluth-Putvinski operator significantly overestimates the avalanche growth rate (points below 1 on the vertical axis) are present at all temperatures. The operator in \Eq{eq:Ch} is thus of general interest. 

Since the electric field spike responsible for the acceleration of runaways during a tokamak disruption is induced by the temperature drop, and therefore occurs slightly later than the drop itself, the temperature is low during the majority of the acceleration process. For significant runaway acceleration, $E/E_{\t{c}}\! \gg\! 1$ is therefore required, and during the initial part of the acceleration process, parameters are likely those corresponding to the blue region of \Fig{fig:K_ratio}b, where the improved avalanche source produces a significantly higher growth rate than the Rosenbluth-Putvinski operator. Post-thermal-quench temperatures in ITER are expected be as low as 10 eV and peak electric fields in disruptions can reach 80 V/m or more \cite{Papp}. Towards the end of the thermal quench, the normalized electric field is then $E/E_{\t{c}}\! \approx\! 1300$ (with $E\!=\!80\,$V/m, $T\!=\!50\,$eV and $n\!=\!1\power{20}\,$m$^{-3}$). A typical ITER disruption would thus (at least initially) be firmly in the blue region of \Fig{fig:K_ratio}b, and the avalanche growth should be significantly higher than what the Rosenbluth-Putvinski source predicts. As the temperature is low, the runaways will also spend a comparatively long time at low momenta ($p\! \ll\! 1$), where the disagreement between the operators is most pronounced. \blue{Note that, according to the figure, an \emph{average} runaway energy of several MeV ($p> 5-10$) is needed for the difference between the growth rates to become small for all $E/E_{\t{c}}$, at which point the most energetic electrons will have reached energies of several tens of MeV or more.} However, since the electric field changes rapidly, the runaways may experience parameters corresponding to both the orange and blue regions in \Fig{fig:K_ratio}b \blue{before reaching such energies}. Further work is therefore needed to assess the overall impact on the avalanche growth of using the improved operator (\ref{eq:Ch}), although it is clear that its use is essential for accurate analysis.

\section{Summary}
Runaway electrons are intimately linked to dynamic scenarios, as they predominantly occur during disruptions and sawtooth events in tokamaks. An accurate description of their dynamics in such scenarios requires kinetic modelling of rapidly changing plasma conditions, and mechanisms such as hot-tail runaway generation add to the already interesting set of phenomena of importance to the evolution of the runaway population. 

In this paper we have described the modelling of several such processes, using the numerical tool \CODE{} to calculate the momentum-space distribution of runaway electrons. In particular, we have investigated rapid-cooling scenarios where hot-tail runaway-electron generation is dominant. Good agreement with previous theoretical work was observed, but \CODE{} simulations also allow for flexible study of a variety of parameter regimes not readily accessible in analytical treatments, and involving other processes such as avalanche generation or synchrotron radiation. 

Furthermore, the full linearized non-relativistic Fokker-Planck-Landau collision operator was discussed, and its implementation described. The operator was found to reproduce the expected Spitzer conductivity in the relevant parameter regime and showed excellent conservation properties. The use of such an operator is essential for the correct current evolution in self-consistent modelling, and in particular when studying the interplay between current and electric-field evolution and runaway-electron generation during a disruption. 

The process of avalanche multiplication of the runaway population via close Coulomb collisions was also considered, and an improved operator, relaxing some of the approximations of the commonly used Rosenbluth-Putvinski operator, was discussed. It was found that the avalanche growth rate can be significantly affected -- increased for low temperatures and high $E/E_{\t{c}}$ and decreased for low $E/E_{\t{c}}$ -- by the use of the new operator. The change to the growth rate can be especially large during the early stages of the runaway acceleration process, thus potentially affecting the likelihood of a given runaway seed transforming into a serious runaway beam, and use of the improved operator is of particular relevance in disruption scenarios. 

The work presented in this paper paves the way for a better understanding of runaway-electron dynamics in rapidly changing scenarios, for instance during tokamak disruptions. It enables more accurate assessment of the risks posed by runaway electrons in situations of experimental interest, particularly in view of future tokamaks such as ITER.

\ack
The authors are grateful to I. Pusztai and E. Hirvijoki for fruitful discussions. This work has been carried out within the framework of the EUROfusion Consortium and has received funding from the Euratom research and training programme 2014-2018 under grant agreement No 633053. The views and opinions expressed herein do not necessarily reflect those of the European Commission. The authors also acknowledge support from Vetenskapsr\aa det, the Knut and Alice Wallenberg Foundation and the European Research Council (ERC-2014-CoG grant 647121).


\end{document}